# Adversarial Training-Aided Time-Varying Channel Prediction for TDD/FDD Systems

Zhen Zhang[1], Yuxiang Zhang[1], Jianhua Zhang[1,*], Feifei Gao[2]

[1] State Key Lab of Networking and Switching Technology, Beijing University of Posts and Telecommunications, Beijing 100876, China
[2] Institute for Artificial Intelligence, Tsinghua University, Beijing 100084, China
* The corresponding author, email: jhzhang@bupt.edu.cn

**Abstract:** In this paper, a time-varying channel prediction method based on conditional generative adversarial network (CPcGAN) is proposed for time division duplexing/frequency division duplexing (TDD/FDD) systems. CPcGAN utilizes a discriminator to calculate the divergence between the predicted downlink channel state information (CSI) and the real sample distributions under a conditional constraint that is previous uplink CSI. The generator of CPcGAN learns the function relationship between the conditional constraint and the predicted downlink CSI and reduces the divergence between predicted CSI and real CSI. The capability of CPcGAN fitting data distribution can capture the time-varying and multipath characteristics of the channel well. Considering the propagation characteristics of real channel, we further develop a channel prediction error indicator to determine whether the generator reaches the best state. Simulations show that the CPcGAN can obtain higher prediction accuracy and lower system bit error rate than the existing methods under the same user speeds.
**Key words:** channel prediction; time-varying channel; conditional generative adversarial network; multipath channel; deep learning.

## I. INTRODUCTION

Accurate downlink channel state information (DL-CSI) is vital to realize coherent detection, precoding, and other various techniques in current and future sixth generation (6G) communication system [1][2]. In the frequency division duplexing (FDD) system, the mobile station (MS) estimates the DL-CSI and feeds it back to the base station (BS). In the time division duplexing (TDD) system, DL-CSI can be estimated by uplink channel state information (UL-CSI) through channel reciprocity. However, in the mobile scenario, channel fading will fluctuate drastically in a short time period [3]. Due to calculation and feedback delay, the channel state information (CSI) is often out of date, which seriously affects the communication system's performance [4].

Motivated by this, many deep learning-based channel prediction techniques are proposed to increase CSI accuracy and reduce the uplink feedback overhead, which combats the fast time-varying channel [5]-[9]. In [5], the authors propose a complex-valued three-dimensional convolutional neural network (CV-3DCNN) to deal with the complex CSI matrices. In [6], the authors develop two deep learning-aided channel estimation methods in multi-cell massive multiple-input-multiple-output (MIMO) systems. In [7]-[9], channel prediction with deep learning can consider the time and frequency domains and does not need the assumption of channel propagation characteristics. However, these researches generally utilize the mean square error (MSE) of the CSI as the loss function that often makes the generated CSI unable to have the same channel characteristics as the real CSI. For example, the relative relationship between multipath cannot be preserved accurately. Hence, a suitable loss function needs to be designed to accurately realize high-dimensional data prediction, which is a thorny problem and generally requires expert knowledge [10].

In [11]-[14], the authors utilize the generative adversarial network (GAN) to capture channel data distribution characteristics with high fitting accuracy. GAN is suitable for a task that requires different kinds of loss functions [15]. In [7], a boundary equilibrium GAN (BEGAN)-based channel prediction method is proposed. However, the BEGAN-based DL-CSI prediction method is time-consuming in real-time prediction and hard to apply to the time-varying channel. Hence, it is necessary to conduct further research on the GAN-based channel prediction method comprehensively.

In this paper, we design a channel prediction method based on the conditional GAN (CPcGAN) for TDD/FDD systems. We give the metrics representing the channel parameter and channel matrix to show the disadvantage of CSI's MSE as the loss function. A channel prediction error (CPError) indicator is developed to make the proposed method consider the local and global characteristics of the CSI matrix comprehensively. Compared with the existing GAN-based channel prediction methods, the proposed method reduces computational complexity significantly and can be applied to time-varying channel prediction. Simulations show that the proposed channel prediction method has significant advantages in CSI's local and global prediction accuracy, leading to a lower bit error rate (BER) than the existing methods.

This paper is organized as follows. The signal and channel model are introduced in Section II. In Section III, we detail our proposed method and metrics. Simulation setup and results are given in Section IV. The conclusions are summarized in Section V.

## II. SYSTEM MODEL

### 2.1 Signal Model

Firstly, let us consider one transmit-receive antenna pair to predict the DL-CSI in an orthogonal frequency division multiplexing (OFDM) system. Then, we extend the method to the MIMO case. The frequency-domain downlink signal (after removing cyclic prefix and performing discrete Fourier transform (DFT)) received at the MS can be expressed as

$$Y_{DL}(t_{DL}, k_{DL}) = H_{DL}(t_{DL}, k_{DL})S_{DL}(t_{DL}, k_{DL}) + W_{DL}(t_{DL}, k_{DL}), \quad (1)$$

where $S_{DL}(t_{DL}, k_{DL})$ and $W_{DL}(t_{DL}, k_{DL})$ are the DFT of transmitted signal and Gaussian noise, respectively. $t_{DL}$ and $k_{DL}$ are the indicators of OFDM symbol and the subcarrier frequency, respectively. $H_{DL}(t_{DL}, k_{DL})$ is the frequency-domain downlink channel. Similarly, the frequency-domain uplink signal received at the BS can be expressed as

$$Y_{UL}(t_{UL}, k_{UL}) = H_{UL}(t_{UL}, k_{UL})S_{UL}(t_{UL}, k_{UL}) + W_{UL}(t_{UL}, k_{UL}). \quad (2)$$

The difference between TDD and FDD systems is whether the uplink and downlink frequency bands are the same.

### 2.2 Channel Model

The geometric channel model is adopted in this paper. Hence, the downlink channel at time $t_{DL}$ is modeled as

$$\boldsymbol{h}_{DL}(t_{DL}, \tau) = \sum_{l=1}^{L(t_{DL})} \alpha_l(t_{DL})e^{j2\pi v_{DL}^l t_{DL}} \times \delta(\tau - \tau_l(t_{DL})), \quad (3)$$

where $L(t_{DL})$ is the number of downlink channel paths. $a_l(t_{DL})$, $v_{DL}^l$, $\tau_l(t_{DL})$ are the downlink path gain, doppler frequency shift, delay, of the $l$th channel path, respectively. The uplink channel at time $t_{UL}$ is modeled as

$$\boldsymbol{h}_{UL}(t_{UL}, \tau) = \sum_{l=1}^{L(t_{UL})} \alpha_l(t_{UL})e^{j2\pi v_{UL}^l t_{UL}} \times \delta(\tau - \tau_l(t_{UL})). \quad (4)$$

At the same time $t$, the uplink and downlink channels have the same number of multipath because the MS and BS positions are reciprocity. Besides, parameters such as path delay and absolute path gain are also frequency-independent [16]. At the different time $t_{UL}$ and $t_{DL}$, due to the spatial consistency of channel, the multipath parameters change continuously over time [17]. Therefore, the channel has a strong correlation in both time and frequency dimensions. The DFT of the downlink channel impulse response for $\tau$ is expressed as (uplink channel is similar)

$$\boldsymbol{H}_{DL}(t, k) = \sum_{l=1}^{L(t)} \alpha_l(t)e^{j2\pi v_{DL}^l t}e^{\frac{-j2\pi k \tau_l(t)}{K}},$$
$$k = 0, 1, \dots K - 1. \quad (5)$$

To study the prediction of multi-frequency time-varying channels, we need channel data for a little consecutive time. In this paper, we utilize the DL-CSI $\boldsymbol{H}_{DL}(t, k) \in \mathbb{C}^{K_{DL} \times T_{DL}}$ to represents the downlink channel of $K_{DL}$ subcarriers and $T_{DL}$ OFDM symbols. Similarly, the UL-CSI $\boldsymbol{H}_{UL}(t, k) \in \mathbb{C}^{K_{UL} \times T_{UL}}$ represents the uplink channel of $K_{UL}$ subcarriers and $T_{UL}$ OFDM symbols. In our work, the $\boldsymbol{H}_{UL}(t, k)$ of the first $T_{UL}$ OFDM symbols is applied to predict the $\boldsymbol{H}_{DL}(t, k)$ of the following $T_{DL}$ OFDM symbols.

## III. ADVERSARIAL TRAINING-AIDED CHANNEL PREDICTION

In conventional channel prediction works, the channel parameter-based prediction methods are tedious and need to be re-estimated iteratively, leading to high computational complexity [18][19]. Deep learning has been widely used in wireless channel research in recent years, and its superiority has been proven [20]-[23]. The existing methods based on deep learning consider the prediction accuracy based on MSE of $\boldsymbol{H}_{DL}(t, f)$ [7]-[9]. However, the neural network produces a blurred result when the loss function aims to minimize the Euclidean distance between the real value and the predicted value [24][25]. The blurred result is that the neural network averages all possible output values to reduce the Euclidean distance, which may lose the relative relationship between the multipath in the channel prediction.

In this work, we design a prediction method that is trained offline. The computational complexity is significantly reduced during the prediction. In the prediction result, we consider the prediction accuracy of the channel matrix $\boldsymbol{H}_{DL}(t, k)$ and the relative relationship between the channel multipath. Hence, the proposed method can obtain high prediction accuracy in both the channel matrix and the channel parameters. In the communication process, we assume that the BS can acquire $\boldsymbol{H}_{UL}(t, k)$ by means of uplink pilot transmitted by MS. Therefore, the proposed method utilizes current time $\boldsymbol{H}_{UL}(t, k)$ to predict the following time $\boldsymbol{H}_{DL}(t, k)$. The proposed method is illustrated in details as follows:

### 3.1 Channel Prediction Based on Conditional GAN

The conditional GAN (cGAN) is a generative model with a conditional constraint. The cGAN's goal is to fit the conditional distribution of the real data rather than utilize a specific parameter as the loss function. Hence, cGAN can capture the statistical distribution of samples and fit the constraint relationship between input and output data. Due to the similar propagation environment, $\boldsymbol{H}_{UL}(t,k)$ and $\boldsymbol{H}_{DL}(t,k)$ are highly relevant in a certain range of time and frequency [26]-[28]. Therefore, the $\boldsymbol{H}_{DL}(t,k)$ prediction problem can be regarded as a channel generation problem that takes $\boldsymbol{H}_{UL}(t,k)$ as a condition to generates the corresponding $\boldsymbol{H}_{DL}(t,k)$.

We design a channel prediction method aided by the cGAN for TDD/FDD systems called CPcGAN. For the CPcGAN, the input of the generator network is the current time real $\boldsymbol{H}_{UL}(t,k)$. The $\boldsymbol{H}_{UL}(t,k)$ is used to predict the following time $\boldsymbol{H}_{DL}(t,k)$. The input of the discriminator network is a data pair composed of $\boldsymbol{H}_{UL}(t,k)$ and $\boldsymbol{H}_{DL}(t,k)$. Here $\boldsymbol{H}_{DL}(t,k)$ may be the real data in the existing communication system, or it may be the fake data generated by the generator. The discriminator judges the matching relationship between the input $\boldsymbol{H}_{UL}(t,k)$ and $\boldsymbol{H}_{DL}(t,k)$ and whether the generated $\boldsymbol{H}_{DL}(t,k)$ is sufficiently consistent with the real data distribution. The proposed CPcGAN framework is shown in Fig. 1.

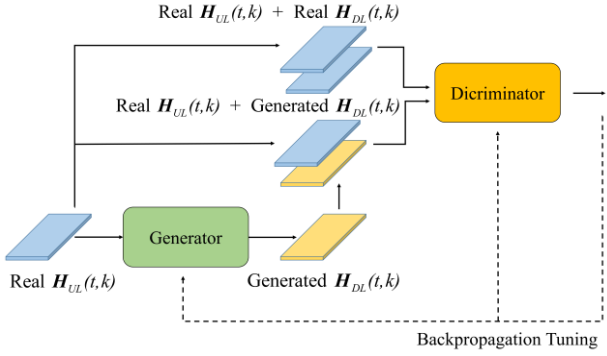

Fig. 1: CPcGAN framework.

The objective of the proposed CPcGAN method can be expressed as

$$L(G,D) = \mathbb{E}[\log D(\boldsymbol{H}_{UL}(t,k),\boldsymbol{H}_{DL}(t,k))] \\ + \mathbb{E}\left[\log\left(1 - D(\boldsymbol{H}_{UL}(t,k),G(\boldsymbol{H}_{UL}(t,k)))\right)\right], \quad (6)$$

where $G$ and $D$ are the generator network and discriminator network, respectively. $\mathbb{E}[\cdot]$ denotes expectation operation. The goal of the discriminator is to maximize the result of (6). Therefore, the discriminator is used to measure the divergence of the two different data pairs distribution. To consider the key channel propagation characteristics and avoid fitting arbitrary noise data, the L1 distance is added to the objective function.

$$R_{L1}(G) = \mathbb{E}\left[\|\boldsymbol{H}_{DL}(t,k) - G(\boldsymbol{H}_{UL}(t,k))\|_1\right]. \quad (7)$$

Hence, the overall objective function of the proposed method is given by

$$L_O(G,D) = \arg\min_G \max_D L(G,D) + \lambda_1 R_{L1}(G), \quad (8)$$

where $\lambda_1$ is the hyperparameter of the model. The optimization goal of the generator $G$ from minimizing (8) is modified to minimizing (9) to avoid saturation of training. The generator aims to reduce the divergence between the generated data and the real data distributions under conditional constraint.

$$L_M(G) = -\mathbb{E}\left[\log\left(D(\boldsymbol{H}_{UL}(t,k),G(\boldsymbol{H}_{UL}(t,k)))\right)\right] \\ + \lambda_1 R_{L1}(G). \quad (9)$$

### 3.2 Metric and Channel Prediction Error

In the existing channel prediction method based on deep learning, the metric is the normalized mean square error (NMSE) of the $\boldsymbol{H}_{DL}(t,k)$ (NMSE$_H$) [7][8], which is given by (10). $\widehat{\boldsymbol{H}}_{DL}(t,k)$ is the predicted value of $\boldsymbol{H}_{DL}(t,k)$. Extended Vehicular A (EVA) model and 5G New Radio (NR) model have the least and the largest number of paths in this paper. The details of datasets are introduced in subsection 4.1. Fig. 2 shows that CSI constantly fluctuates around a specific median range in the entire frequency dimension. In other words, the fluctuation of CSI is slight in the relatively close time and frequency range. The problem is similar to the situation in computer vision where the difference between adjacent pixel values is small and sharp edges are not visible, which may cause the neural network to produce average results and lose the propagation law of channel. Since channel propagation is dominated by a limited number of multipath, the channel multipath delay is considered to characterize the main characteristics of the channel.

$$\text{NMSE}_H = \mathbb{E}\left[\frac{\|\boldsymbol{H}_{DL}(t,k) - \widehat{\boldsymbol{H}}_{DL}(t,k)\|^2}{\|\boldsymbol{H}_{DL}(t,k)\|^2}\right]. \quad (10)$$

In Fig. 3, the CSI is transformed into the delay dimension with only limited multipath. Therefore, a slight multipath offset can cause a large error. To further amplify the multipath effect, the power delay profile (PDP) is used for reference. It can be seen in Fig. 4 that when the PDP is used to characterize the channel, the channel has a sharp boundary, which will help the neural network to generate a more realistic wireless channel. We need a new metric to represent the multipath characteristics of the channel. Hence, the time-varying PDP (TVPDP) $\boldsymbol{P}(t,\tau)$ that reflects the channel multipath relative relationship in the $T_{DL}$ OFDM symbols period is defined in (11)-(13).

$$\boldsymbol{h}(t,\tau) = \frac{1}{K}\sum_{k=0}^{K-1}\boldsymbol{H}_{DL}(t,k)W_K^{-\tau k}, \\ \tau = 0,1,\ldots K-1. \quad (11)$$

$$W_K = e^{\frac{-j2\pi}{K}}. \quad (12)$$

$$\boldsymbol{P}(t,\tau) = 10 \times \lg(|\boldsymbol{h}(t,\tau)|^2). \quad (13)$$

The proposed metric--NMSE of $\boldsymbol{P}(t,\tau)$ (NMSE$_P$) is defined as

$$\text{NMSE}_P = \mathbb{E}\left[\frac{\|\boldsymbol{P}(t,\tau) - \widehat{\boldsymbol{P}}(t,\tau)\|^2}{\|\boldsymbol{P}(t,\tau)\|^2}\right], \quad (14)$$

where $\widehat{\boldsymbol{P}}(t,\tau)$ and $\boldsymbol{P}(t,\tau)$ are the TVPDP calculated based on the $\widehat{\boldsymbol{H}}_{DL}(t,k)$ and $\boldsymbol{H}_{DL}(t,k)$, respectively.

It is worth noting that minimizing a single metric does not mean the prediction result is the best. The nonlinear mapping of the logarithmic function makes the $\text{NMSE}_\text{H}$ and $\text{NMSE}_\text{P}$ reflect different error dimensions. The $\text{NMSE}_\text{H}$ is utilized to measure the size of the DL-CSI's error in the frequency dimension, and the $\text{NMSE}_\text{P}$ is utilized to measure the relative error between multipath in the delay dimension. Therefore, the accurate channel prediction result is achieved only when both $\text{NMSE}_\text{H}$ and $\text{NMSE}_\text{P}$ are relatively small.

To comprehensively reflect the channel's local and global characteristics, we develop a new indicator that focuses on channel parameters and channel matrix based on the original framework of the proposed method. The indicator called channel prediction error (CPError) is defined in (15) to judge the convergence effect of the CPcGAN model.

$$\text{CPError} = \text{NMSE}_\text{H} + \lambda_2 \times \text{NMSE}_\text{P}, \quad (15)$$

where $\lambda_2$ is also a hyperparameter of the CPcGAN. The $\text{NMSE}_\text{P}$ can maintain the relative relationship between multipath so that local features can be preserved. The $\text{NMSE}_\text{H}$ can retain the global characteristics and phase information of the channel. Therefore, the CPError can make the generated CSI achieve the best in channel amplitude and phase. We utilize CPError to determine whether model training is completed without changing the CPcGAN's training way. The main steps of the proposed method are described in Algorithm 1.

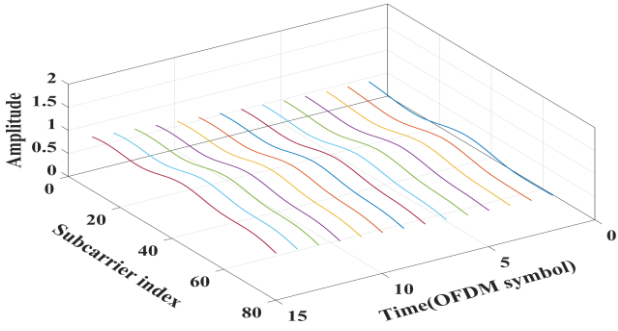 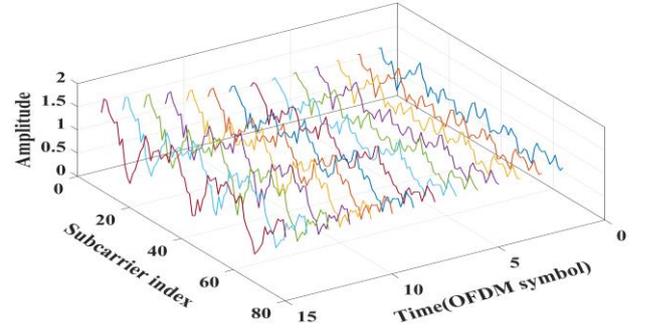

(a) EVA channel with user speed 50 km/h.  (b) 5G NR channel with user speed 300 km/h.

Fig. 2: CSI's frequency dimension.

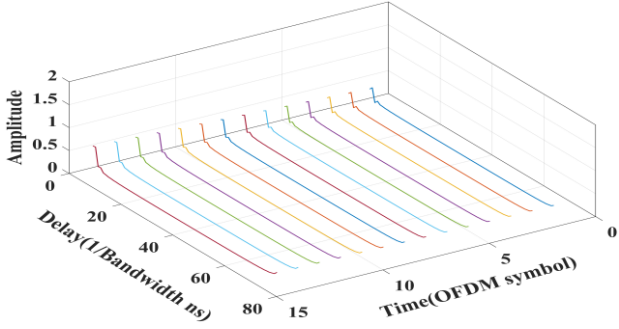 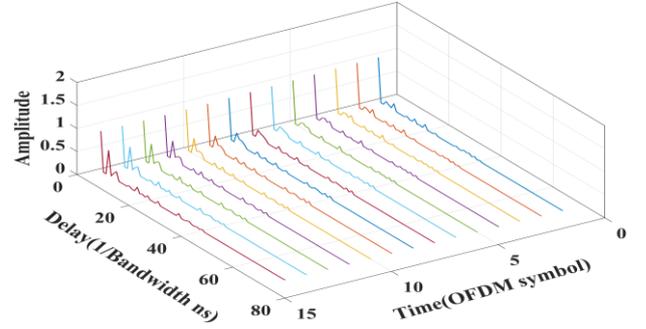

(a) EVA channel with user speed 50 km/h.  (b) 5G NR channel with user speed 300 km/h.

Fig. 3: CSI's multipath delay dimension.

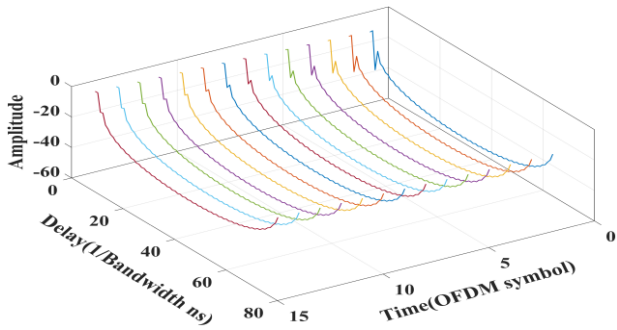 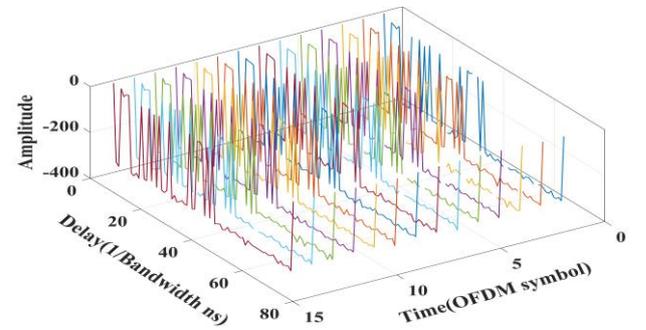

(a) EVA channel with user speed 50 km/h.  (b) 5G NR channel with user speed 300 km/h.

Fig. 4: CSI's PDP.

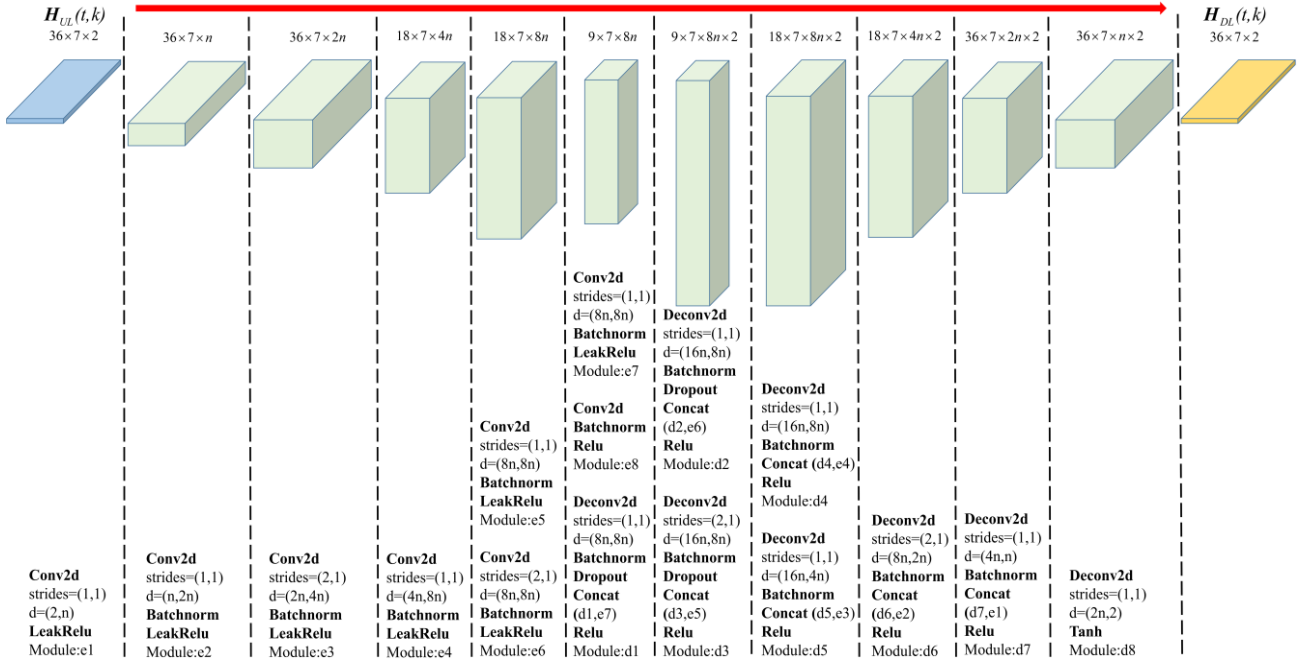

Fig. 5: Generator.

**Algorithm 1** CPcGAN

Train:
1 : Initialize all variables and network structure of the generator and discriminator models;
2 : for i = 1 to Epoch:
3 :　Randomly sort the training set data and group them according to the batch size;
4 :　for idx = 1 to Number of batch groups:
5 :　　Calculate $\widehat{\boldsymbol{H}}_{DL}(t,k) = G(\boldsymbol{H}_{UL}(t,k))$;
6 :　　Calculate discriminator loss then update discriminator parameters:
$$L(D) = \mathbb{E}[\log D(\boldsymbol{H}_{UL}(t,k), \boldsymbol{H}_{DL}(t,k))]$$
$$+\mathbb{E}\left[\log\left(1 - D(\boldsymbol{H}_{UL}(t,k), \widehat{\boldsymbol{H}}_{DL}(t,k))\right)\right]$$
7 :　　Calculate generator loss then update generator parameters twice:
$$L_M(G) = -\mathbb{E}\left[\log\left(D(\boldsymbol{H}_{UL}(t,k), G(\boldsymbol{H}_{UL}(t,k)))\right)\right]$$
$$+\lambda_1 R_{L1}(G).$$
8 :　　counter = counter + 1;
9 :　　Calculate CPError with validation set every 100 batches and save model:
　　　if Mod(counter, 100) == 0:
　　　　CPError = $\text{NMSE}_H + \lambda_2 \times \text{NMSE}_P$,
10:　　end if
11:　end for
12: end for
13: Determine the best generator by CPError.

Test:
1 : Use the generator network to predict DL-CSI:
$$\widehat{\boldsymbol{H}}_{DL}(t,k) = G(\boldsymbol{H}_{UL}(t,k)).$$

## IV. SIMULATION SETUP AND RESULTS

In this section, we first introduce the datasets and network structure. Then, we compare the simulation results of the proposed method and the existing method at the same speed and show superiority of the proposed method. Finally, we offer the expandability and generalization results of the proposed method in terms of multiple antennas, high-speed mobile, and various channels.

### 4.1 Dataset Setup

All channel datasets are generated by the Vienna LTE-A Downlink link-level simulator [29] and the Vienna 5G link-level simulator [30]. Three fading channel models, EVA model, Extended Typical Urban (ETU) model, and 5G NR channel model based on the 3GPP 38.901 [17], are utilized to evaluate the performance of the proposed approach. For each $\boldsymbol{H}_{UL}(t,k)$ or $\boldsymbol{H}_{DL}(t,k)$ sample, the size is $36 \times 7$, representing the channel of 36 subcarriers and 7 OFDM symbols. The 36 subcarriers in 7 OFDM symbols are equivalent to 3 physical resource blocks in a slot [31]. We utilize the datasets of the FDD system consistent with [5][6]. In every dataset, there are a total of 40 K data samples. 35 K samples are taken as the training set, 1 K samples as the validation set, and 4 K samples as the test set. The complex CSI matrix is divided into a real part and an imaginary part, and they are regarded as two channels in an image.

### 4.2 Network Structure

The generator is designed according to the encoder-decoder framework and shown in Fig. 5. The encoder-decoder framework is more in line with the channel prediction idea where environmental information around the channel is extracted from $H_{UL}(t,k)$ and mapped to $H_{DL}(t,k)$ [32]. The network is divided into modules and renamed according to the encoding and decoding parts, such as $e_1$, $d_1$. The $e_1$ module includes a convolutional layer and a LeakRelu layer. $d=(2,n)$ means that the input data channels and the convolution kernels are 2 and $n$, respectively. $H_{UL}(t,k)$ generates the $36 \times 7 \times n$ data block in the second column after $e_1$ module computation. Concat $(d_1, e_7)$ means to merge the output of $d_1$ and $e_7$ module.

The structure of the discriminator is shown in Fig. 6. The discriminator is designed to continuously compress the input data and finally extract high-level information. Since the loss function is cross-entropy, the final network needs a fully connected layer to change the output dimension to 1 and the sigmoid activation function to limit the output range.

Adam optimizer is used in the CPcGAN model. The learning rate is set to 0.0002, $\beta_1=0.5$, $\beta_2=0.999$. The kernel size is (3,3), and $n$ is 64. The values of $\lambda_1$, $\lambda_2$ and batch size are 100, 1 and 32, respectively. The real and imaginary parts of CSI are linearly normalized to the interval [-1,1]. In order to avoid the rapid convergence of the discriminator network, the generator network is updated twice as every time the discriminator network update [33].

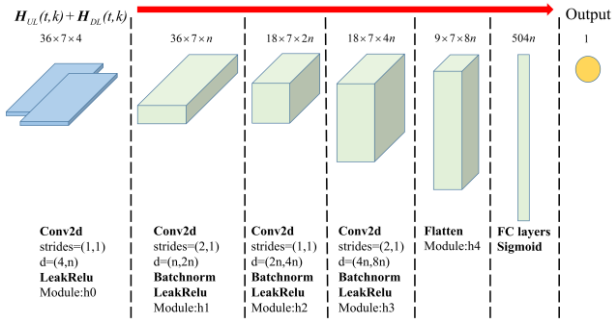

Fig. 6: Discriminator.

## 4.3 Simulation Results Comparison

In this subsection, we compare the proposed method with the linear minimum mean squared error (LMMSE) method [34], convolutional neural network (CNN)-based method, and the BEGAN-based method from metric results and BER two aspects in the single-input-single-output (SISO) channel. In order to introduce the time-varying characteristics of the channel, the user speed is set to 50 km/h.

The LMMSE method predict DL-CSI at target time $t$ and subcarrier frequency $k$ is given by

$$\hat{H}_{LMMSE}(t,k) = R(t,k)H_{UL}, \quad (16)$$

where the $H_{UL}$ is the known UL-CSI. The $\hat{H}_{LMMSE}(t,k)$ is the predicted DL-CSI. The coefficient matrix $R(t,k)$ is obtained by minimizing the MSE of the predicted values [35]. The $R(t,k)$ is defined in (17)-(19).

$$R(t,k) = R_{H_{DL}H_{UL}}(t,k)R_{H_{UL}H_{UL}}^{-1}, \quad (17)$$
$$R_{H_{DL}H_{UL}}(t,k) = \mathbb{E}[H_{DL}(t,k)H_{UL}^H], \quad (18)$$
$$R_{H_{UL}H_{UL}} = \mathbb{E}[H_{UL}H_{UL}^H]. \quad (19)$$

The CNN-based approach utilizes the MSE of $H_{DL}(t,k)$ as the loss function. The BEGAN-based method also introduces the idea of adversarial training. However, the method of image completion is not suitable for time-varying channel prediction. All details of CNN-based and BEGAN-based methods can be found in [7].

### 4.3.1 Metric Results Comparison

The NMSE$_H$ and NMSE$_P$ results of four prediction methods are shown in Table I. In Fig. 7 and Fig. 8, the visualized PDP results and $H_{DL}(t,k)$ results of 6 DL-CSI samples are given, respectively. The PDP result is calculated from the same time slot of corresponding $H_{DL}(t,k)$.

Since CNN aims to reduce the MSE of $H_{DL}(t,k)$, the NMSE$_H$ of the CNN-based method is smaller than the BEGAN-based method in all channel datasets. However, for a CNN whose loss function only considers the CSI's MSE, it is challenging to capture the channel's multipath characteristics accurately. In Fig. 7, the PDP results of the $H_{DL}(t,k)$ predicted by the CNN-based method lose the real channel's multipath relationship. In the EVA and ETU datasets, the NMSE$_P$ of the CNN-based method is the largest among the four methods. In the 5G NR dataset, the NMSE$_P$ of the CNN-based method is eight times bigger than the LMMSE method and the CPcGAN method. Unlike the CNN-based method, the LMMSE method by channel correlation can retain the channel multipath characteristics, so there is a lower NMSE$_P$ in all datasets. The BEGAN-based method introduces adversarial training that aims to generate samples that conform to the real data distribution. However, the BEGAN-based method only utilizes a part of the CSI to select the entire CSI sample. Hence, it is challenging to reduce NMSE$_H$, and the capability of reducing NMSE$_P$ is also limited, which makes it difficult for the BEGAN-based method to achieve relatively high prediction accuracy.

Combining the adversarial training idea and the CPError indicator, the proposed CPcGAN method can get the best results on NMSE$_H$ and NMSE$_P$ among the all methods. In all simulations, the LMMSE method is only superior to the CPcGAN method under the NMSE$_P$ of the ETU dataset. CPcGAN takes $H_{UL}(t,k)$ as a constraint, and it is easier to capture the real channel distribution than other methods. In Fig. 7, the PDP curve acquired by the CPcGAN method matches the PDP curve of the real sample the best compared with the above methods. The relative relationship between multipath is fitted well. Fig. 8 shows the absolute amplitude values of $\boldsymbol{H}_{DL}(t,k)$ predicted by the CPcGAN method for the above sample. The $\boldsymbol{H}_{DL}(t,k)$ generated by the CPcGAN method is very consistent with the real channel propagation characteristic. Besides, the CPcGAN method no longer repeats the gradient descent computation like the BEGAN-based method during prediction and can achieve similar computational complexity as the CNN-based method.

TABLE I: Metric results comparison.

| Datasets | Metrics | LMMSE | CNN | BEGAN | CPcGAN |
|---|---|---|---|---|---|
| EVA | $NMSE_H$ | 0.0454 | 0.0214 | 0.0877 | 0.0170 |
| | $NMSE_P$ | 0.0063 | 0.0237 | 0.0201 | 0.0063 |
| ETU | $NMSE_H$ | 0.0452 | 0.0518 | 0.0830 | 0.0164 |
| | $NMSE_P$ | 0.0051 | 0.0404 | 0.0211 | 0.0076 |
| 5G NR | $NMSE_H$ | 0.0104 | 0.0839 | 0.1405 | 0.0037 |
| | $NMSE_P$ | 0.0050 | 0.0427 | 0.0696 | 0.0046 |

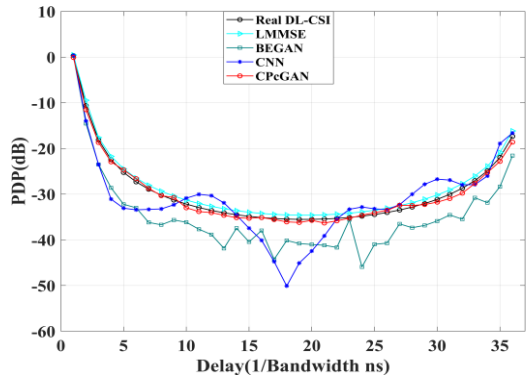

(a) EVA DL-CSI 1's PDP.

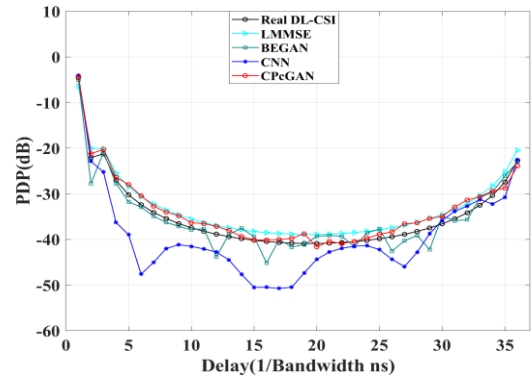

(b) EVA DL-CSI 2's PDP.

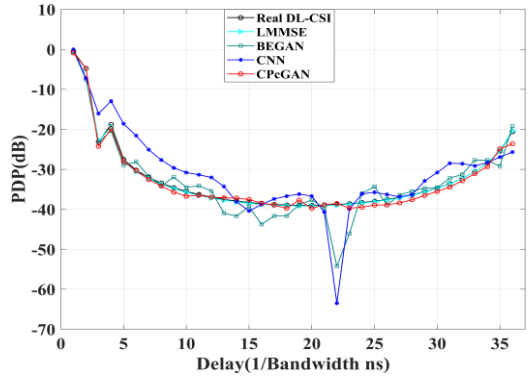

(c) ETU DL-CSI 1's PDP.

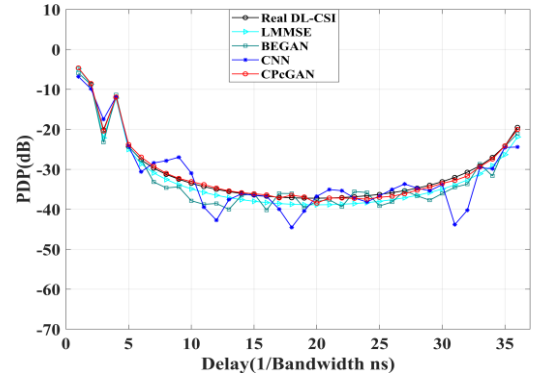

(d) ETU DL-CSI 2's PDP.

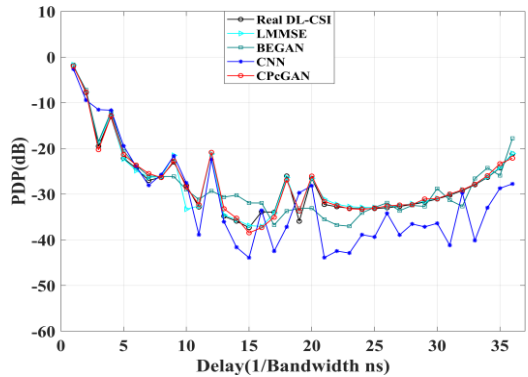

(e) 5G NR DL-CSI 1's PDP.

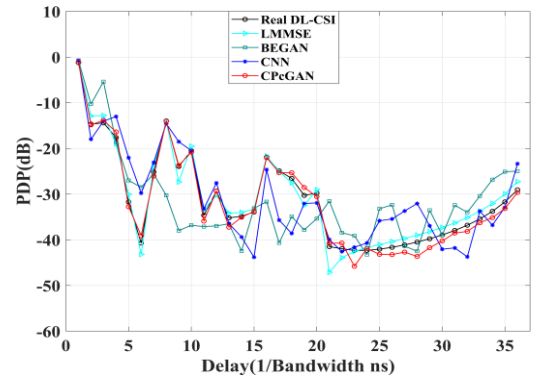

(f) 5G NR DL-CSI 2's PDP.

Fig. 7: The visualized PDP results of the third time slot of the 6 different DL-CSI samples.

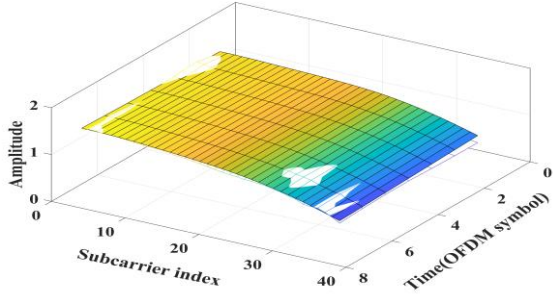
(a) EVA DL-CSI 1's absolute amplitude.

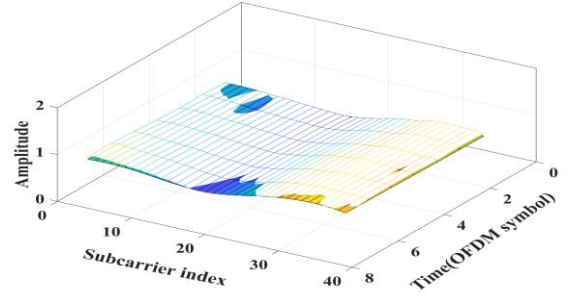
(b) EVA DL-CSI 2's absolute amplitude.

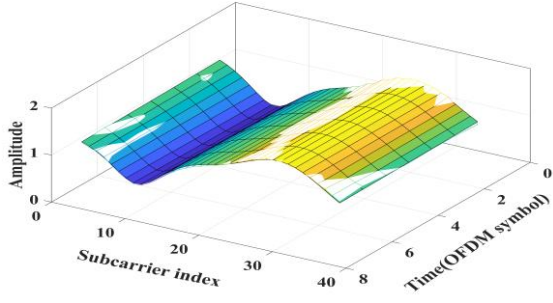
(c) ETU DL-CSI 1's absolute amplitude.

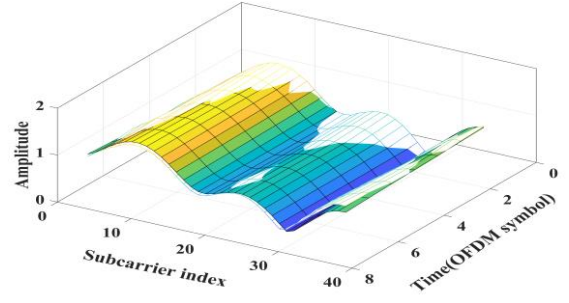
(d) ETU DL-CSI 2's absolute amplitude.

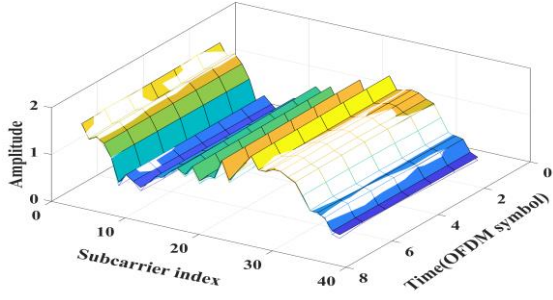
(e) 5G NR DL-CSI 1's absolute amplitude.

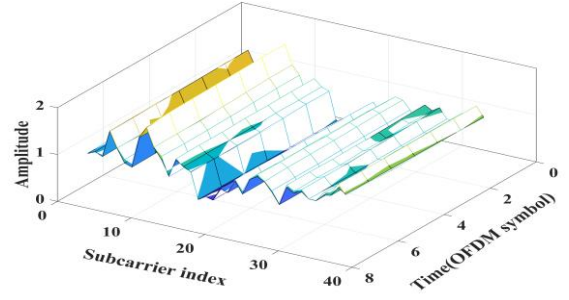
(f) 5G NR DL-CSI 2's absolute amplitude.

Fig. 8: The visualized $H_{DL}(t,k)$ prediction results of the 6 samples above. The real $H_{DL}(t,k)$ is depicted as a colored curved surface. The predicted $H_{DL}(t,k)$ is depicted as a mesh grid plot.

#### 4.3.2 BER Comparison

BER can show the prediction accuracy of the four prediction methods under time-varying channels directly. Therefore, an OFDM communication system is simulated to compare the BER performance. The predicted $H_{DL}(t,k)$ is used at the BS only to perform the pre-equalization operation. The quadrature phase shift keying (QPSK) modulation is performed. Fig. 9- Fig. 11 show the BER curves of the four different channel prediction methods under the three time-varying channels.

The $NMSE_H$ and $NMSE_P$ of the BEGAN-based method are the highest compared with other methods, which means the worst prediction accuracy. Therefore, the BER is the highest among the four methods. CPcGAN learns the real CSI data distribution and relies on CPError to determine the best channel parameters and channel matrix balance point. Hence, the CPcGAN can capture the time-varying channel characteristics comprehensively. The BER of the CPcGAN method is always lower than the CNN-based method and LMMSE method. For more accurate and complex 5G NR channels, the CPcGAN method shows better BER performance. When the SNR is 20 dB, the BER of the CPcGAN method is 53.9%, 82.5%, 96.3% lower than the CNN-based method under three different channel conditions, respectively. Similarly, the BER of the CPcGAN method is 78.5%, 77.7%, 63.6% lower than the LMMSE method under three different channel conditions, respectively.

### 4.4 Further Discussion

In this subsection, the expandability and generalization of the proposed method will be further discussed. The simulation results are given mainly for MIMO extension, various user speeds, and various channel models, showing the expandability and generalization of proposed method.

#### 4.4.1 MIMO Extension

In order to verify the proposed method is suitable for MIMO channels, the proposed method is tested under the conditions of $2 \times 2$ MIMO EVA channel and 5G NR channel (with the least and the largest number of multipath, respectively).

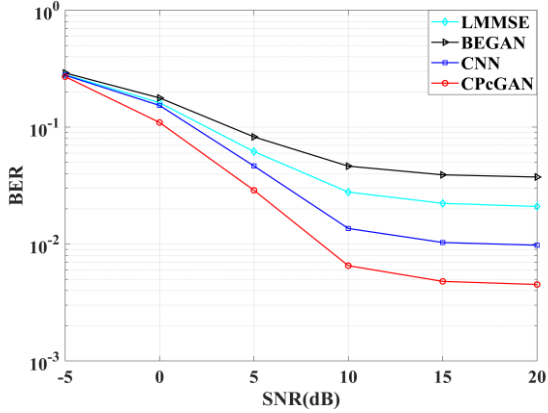

Fig. 9: EVA BER comparison.

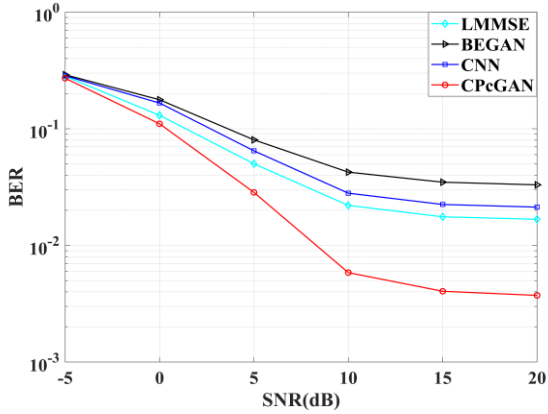

Fig. 10: ETU BER comparison.

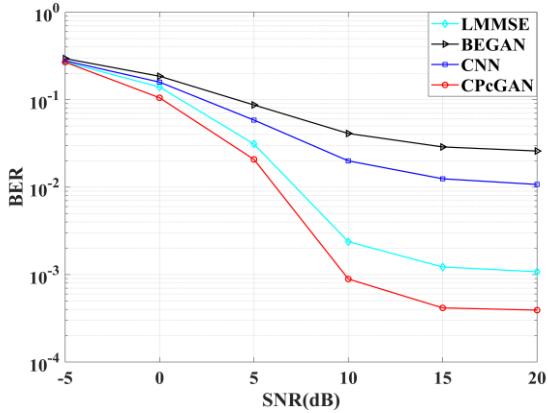

Fig. 11: 5G NR BER comparison.

For a $N_R \times N_T$ MIMO channel (with $N_R$ receive antennas and $N_T$ transmit antennas), we treat it as $N_R \times N_T$ Tx-Rx links. The current time UL-CSI of each Tx-Rx link is fed to the generator of the CPcGAN method, and the generator predicts the following time DL-CSI result of each Tx-Rx link. To verify the expandability of the proposed method, it is worth noting that our training data is still the SISO channel datasets and the test data utilizes the corresponding MIMO channel datasets. In other words, the MIMO channel verification model in this subsection is precisely the same as the model under the SISO channel. Table II shows the DL-CSI prediction average errors of the CPcGAN method under $2 \times 2$ MIMO channels. Fig. 12 and Fig. 13 show the BER results under two different MIMO channels.

TABLE II: MIMO channel metric results comparison.

| Datasets | Metrics | $1 \times 1$ SISO | $2 \times 2$ MIMO |
|---|---|---|---|
| EVA | NMSE$_H$ | 0.0170 | 0.0162 |
|  | NMSE$_P$ | 0.0063 | 0.0073 |
| 5G NR | NMSE$_H$ | 0.0037 | 0.0035 |
|  | NMSE$_P$ | 0.0046 | 0.0044 |

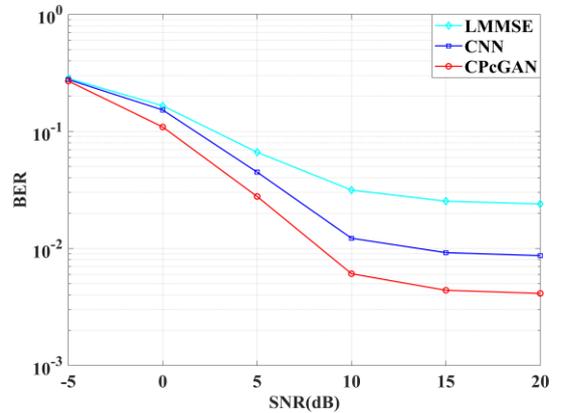

Fig. 12: MIMO EVA BER comparison.

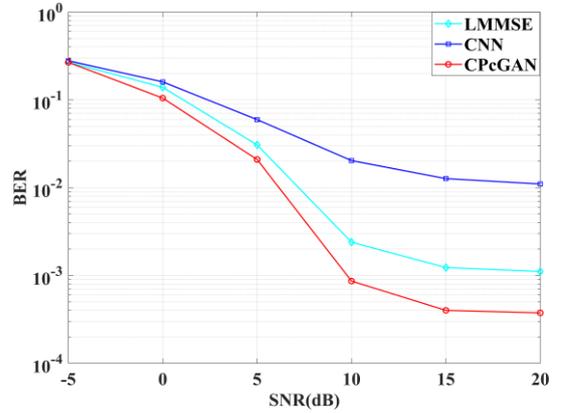

Fig. 13: MIMO 5G NR BER comparison.

Table II shows that the model trained under a single link can be well used to predict the MIMO channel DL-CSI with the corresponding UL-CSI of the MIMO channel as input. Fig. 12 and Fig. 13 show that the proposed method can achieve the best results under MIMO channel conditions compared with other prediction methods. Also, MIMO DL-CSI prediction can obtain a similar BER to the single link channel, meaning that the proposed method is suitable for MIMO channels.

### 4.4.2 Various User Speeds

When the user speed increases, the time-varying channel will become more severe, which may bring challenges to the performance of channel prediction methods. In this subsection, we give the proposed method's NMSE results at various user speeds and explore the performance boundary of the proposed method. Finally, the generalization results of the proposed method at different speeds are given.

We trained the models for EVA and 5G NR channels at different user speeds from 50 km/h to 300 km/h. Fig. 14 and Fig. 15 show the $NMSE_H$ and $NMSE_P$ results under the EVA channel. Fig. 16 and Fig. 17 show the $NMSE_H$ and $NMSE_P$ results under the 5G NR channel. It can be seen from Fig. 14-Fig. 17 that when the user speed increases, the performance of the traditional LMMSE method decreases significantly. Namely, the $NMSE_H$ and $NMSE_P$ both increase rapidly. Although the prediction error of the CNN-based approach and the BEGAN-based approach growth relatively slow, the performance is also poor at high speed. The proposed CPcGAN method is the only way to maintain a small error at each speed among all the simulation methods. For the $NMSE_H$ results, the CPcGAN method at a speed of 300 km/h is still lower than the LMMSE method at 50 km/h.

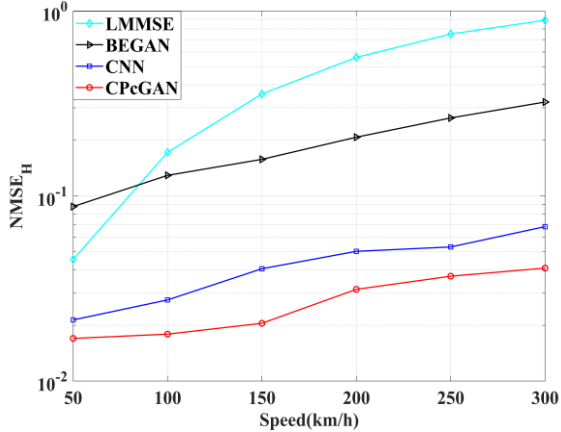

Fig. 14: EVA $NMSE_H$ comparison.

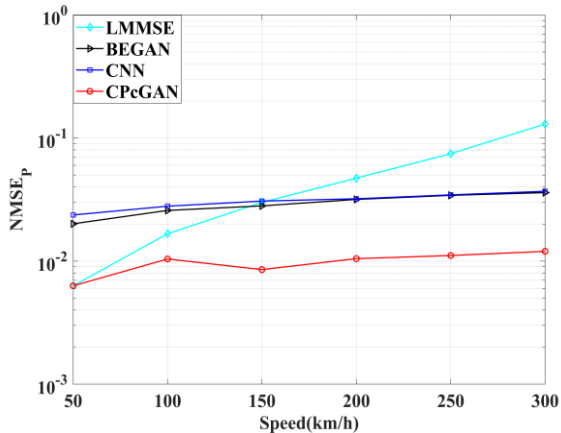

Fig. 15: EVA $NMSE_P$ comparison.

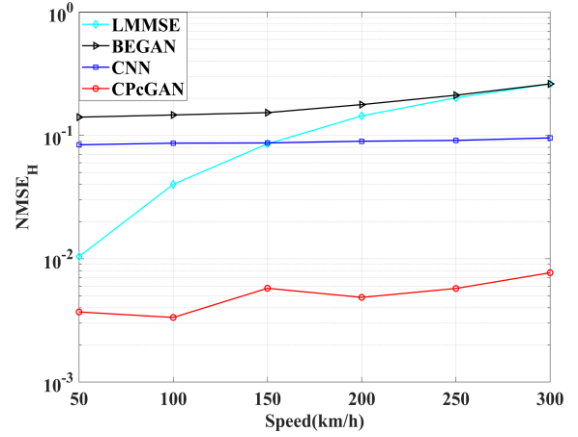

Fig. 16: 5G NR $NMSE_H$ comparison.

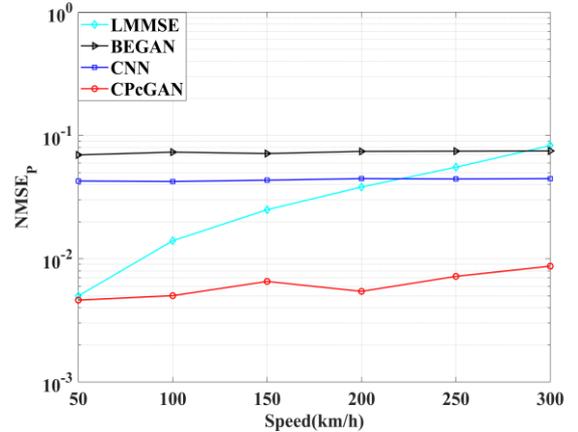

Fig. 17: 5G NR $NMSE_P$ comparison.

The higher user speed has more severe time-varying channel characteristics, so the model trained at 300 km/h can capture more complex time-varying channel characteristics. To study the generalization of the proposed CPcGAN method at different user speeds, we utilize a model trained at 300 km/h to test at different speeds. As shown in Fig. 18, the model trained at higher user speed can maintain stable prediction performance at various other speeds, indicating that the proposed method has good time-varying channel generalization characteristics.

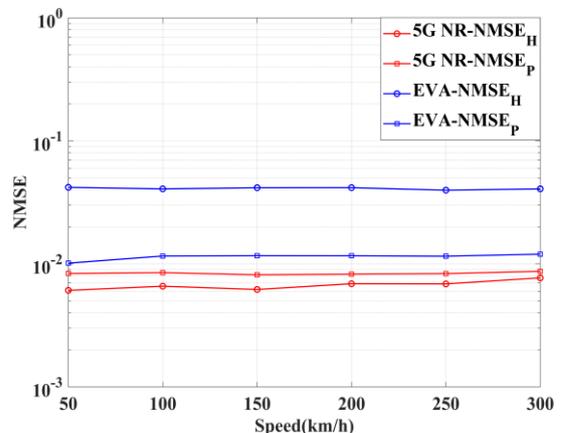

Fig. 18: Generalization test of various user speeds.

### 4.4.3 Various Channel Models

To further study the generalization of the proposed method, a question arises whether a single network can be used for different channel DL-CSI prediction. To answer this question, we merge EVA and 5G NR channel data. We train the generator and discriminator of the CPcGAN method to predict the DL-CSI of two different channels by the combined data. Eventually, the prediction error of proposed method under the combined dataset is similar to or even lower than the LMMSE method. Therefore, when enough data can be obtained, the proposed method can handle different channels concurrently. Fig. 19 shows the DL-CSI results of EVA and 5G NR channels predicted by the LMMSE method. Fig. 20 shows the above same DL-CSI sample prediction results predicted by the CPcGAN method.

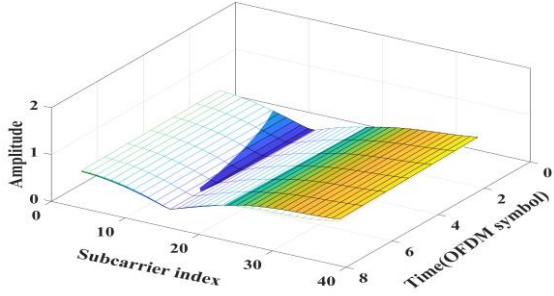 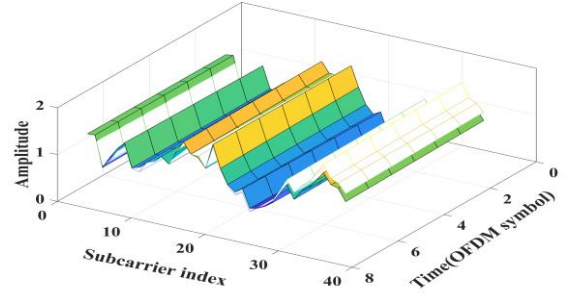

(a) EVA DL-CSI's absolute amplitude.   (b) 5G NR DL-CSI's absolute amplitude.

Fig. 19: The visualized $\boldsymbol{H}_{DL}(t,k)$ prediction results of the 2 samples predicted by the LMMSE method. The real $\boldsymbol{H}_{DL}(t,k)$ is depicted as a colored curved surface. The predicted $\boldsymbol{H}_{DL}(t,k)$ is depicted as a mesh grid plot.

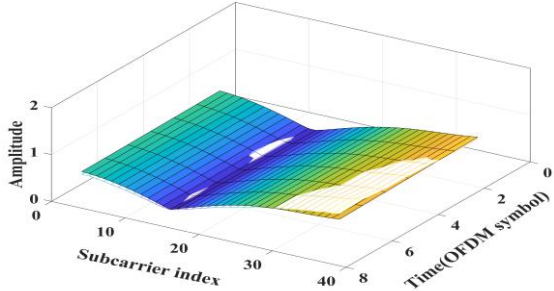 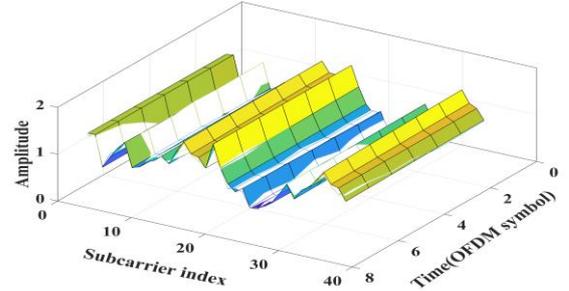

(a) EVA DL-CSI's absolute amplitude.   (b) 5G NR DL-CSI's absolute amplitude.

Fig. 20: The visualized $\boldsymbol{H}_{DL}(t,k)$ prediction results of the above 2 samples predicted by CPcGAN method. The real $\boldsymbol{H}_{DL}(t,k)$ is depicted as a colored curved surface. The predicted $\boldsymbol{H}_{DL}(t,k)$ is depicted as a mesh grid plot.

## V. CONCLUSIONS

We have proposed a novel time-varying channel prediction method based on cGAN to obtain more realistic prediction results. The proposed method is trained offline, which is suitable for the current TDD/FDD systems CSI prediction. Simulations show that the proposed method incorporating more real multipath propagation law achieves higher prediction accuracy than the existing methods. With the continuous increase of user speed, the proposed method can adapt to the changes of fast time-varying channels, and the prediction error is the most stable among the existing methods, demonstrating the advantage of utilizing the adversarial training technique. Considering the difficulty of traditional deep learning loss function design, more wireless communication applications that benefit from the adversarial training technique deserve further exploration.

## ACKNOWLEDGEMENT

This research is supported by Outstanding Youth Fund of National Natural Science Foundation of China (No. 61925102).

## Biographies

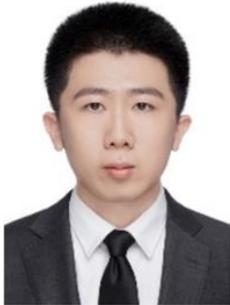

**Zhen Zhang**, received the B.S. degree in communication engineering from the China University of Petroleum (East China), in 2017, where he is currently pursuing the Ph.D. degree in information and communication engineering with the Beijing University of Posts and Telecommunications. His current research interests include channel modeling, OTA testing, machine learning, and so on. Email: zhenzhang@bupt.edu.cn.

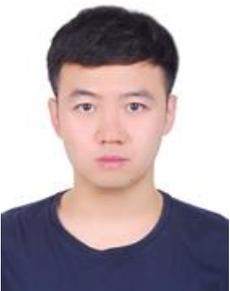

**Yuxiang Zhang**, received the B.S. degree from the Dalian University of Technology, in 2014, and the Ph.D. degree from the Beijing University of Posts and Telecommunications, in 2020, where he is currently a Post Doctoral Fellow. His current research interests include channel modeling, massive MIMO, OTA testing and so on.

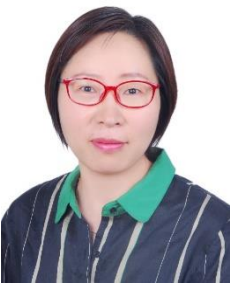

**Jianhua Zhang**, received her Ph.D. degree in circuit and system from Beijing University of Posts and Telecommunication (BUPT) in 2003 and now is professor of BUPT. She has published more than 200 articles in referred journals and conferences and 40 patents. She received several paper awards, including 2019 SCIENCE China Information Hot Paper, 2016 China Comms Best Paper, 2008 JCN Best Paper etc. She received two national novelty awards for her contribution to the research and development of Beyond 3G TDD demo system with 100Mbps@20MHz and 1Gbps@100MHz respectively. She received the second prize for science novelty from Chinese Communication Standards Association for her contributions to ITU-R 4G (ITU-R M.2135) and 3GPP Relay channel model (3GPP 36.814). From 2012 to 2014, she did the 3-dimensional (3D) channel modeling work and contributed to 3GPP 36.873 and is also the member of 3GPP "5G channel model for bands up to 100 GHz". She was the Drafting Group (DG) Chairwoman of ITU-R IMT-2020 channel model and led the drafting of IMT.2412 Channel Model Section. Her current research interests include 5G and 6G, artificial intelligence, data mining, especially in massive MIMO and millimeter wave channel modeling, channel emulator, OTA testing and etc. Webpage: www.zjhlab.net.

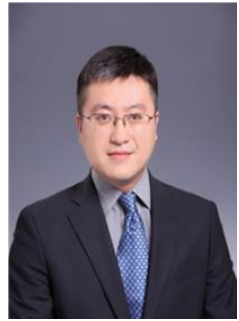

**Feifei Gao**, received the B.Eng. degree from Xi'an Jiaotong University, in 2002, the M.Sc. degree from McMaster University, in 2004, and the Ph.D. degree from National University of Singapore, in 2007. Since 2011, he joined the Department of Automation, Tsinghua University, Beijing, China, where he is currently an Associate Professor. Prof. Gao's research interests include signal processing for communications, array signal processing, convex optimizations, and artificial intelligence assisted communications. He has authored/coauthored more than 120 refereed IEEE journal papers and more than 120 IEEE conference proceeding papers.

Prof. Gao has served as an Editor of IEEE Transactions on Wireless Communications, IEEE Transactions on Cognitive Communications and Networking, IEEE Signal Processing Letters, IEEE Communications Letters, IEEE Wireless Communications Letters, and China Communications. He has also served as the symposium co-chair for 2019 IEEE Conference on Communications (ICC), 2018 IEEE Vehicular Technology Conference Spring (VTC), 2015 IEEE Conference on Communications (ICC), 2014 IEEE Global Communications Conference (GLOBECOM), 2014 IEEE Vehicular Technology Conference Fall (VTC), as well as Technical Committee Members for many other IEEE.